# Design and Implementation of a Microservice-Architecture Master Control System for AIMS

Li-Yue Tong[1,2], Jia-Ben Lin[1,2,3], Jun-Feng Hou[1,2,3], Yuan-Yong Deng[1,2,3], Dong-Guang Wang[1,2,3], Guang-Qian Liu[4], Song-Bo Xu[5], Shang-Jie Ren[5], Lian-Wei Zhao[6], Zhi-Wei Feng[1,2], Wei Duan[1,2], Ming-Fu Shao[1,2,3], Hui Wang[1,2,3] and Chen Yang[1,2,3]

[1] National Astronomical Observatories, Chinese Academy of Sciences, Beijing 100101, China; *jiabenlin@bao.ac.cn*

[2] State Key Laboratory of Solar Activity and Space Weather, National Astronomical Observatories, Chinese Academy of Sciences, Beijing, 100101, China

[3] University of Chinese Academy of Sciences, Beijing 100049, China

[4] Yunnan Observatories, Chinese Academy of Sciences, Kunming 650216, China

[5] Xi'an Institute of Optics and Precision Mechanics, Chinese Academy of Sciences, Xi'an 710119, China

[6] Shanghai Institute of Technical Physics, Chinese Academy of Sciences, Shanghai 200083, China

**Abstract** The mid-infrared solar magnetic field telescope AIMS (An Infrared System for the Accurate Measurement of Solar Magnetic Field) is the first ground-based telescope designed to directly measure solar magnetic fields via Zeeman splitting in the 8–14 $\mu$m band, overcoming the century-long bottleneck of model-dependent indirect measurements. Its remote high-altitude site, heterogeneous multi-institute components, and complex observation modes comprising Fourier Transform Infrared (FTIR) spectropolarimetry and broadband imaging demand a highly autonomous Master Control System (MCS). We present the design and implementation of the AIMS MCS, featuring three key contributions: (1) an L0–L5 telescope automation classification inspired by the SAE J3016 autonomous driving standard, providing well-defined boundaries and a progressive evolution roadmap; (2) a three-layer system framework—device control, autonomy support, and central decision—implemented with a microservice software architecture that achieves loose coupling, high cohesion, and continuous integration of heterogeneous components; and (3) a suite of key enabling technologies including automatic pointing/tracking, autofocus via lucky-frame selection combined with power spectral ratio analysis, and environment-adaptive observation integrating auto-exposure, cloud detection, and power/thermal monitoring. The MCS has been validated across three telescopes at progressive automation levels: AIMS itself (L1, stable operation >2 years since April 2024), the WenQuan Solar Magnetic Field Telescope (L2, stable operation >3 years since February 2023), and the Solar Full-disk Multi-layer Magnetograph (SFMM) (L2 routine operation >2.5 years since October 2023, with L3 autonomous functions verified through dedicated testing). Collectively, these deployments demonstrate the feasibility and stability of the proposed architecture for progressive telescope automation.

## 1 INTRODUCTION

The solar magnetic field is widely regarded as the primary physical quantity governing solar activity (Deng et al. 2020). Since George Ellery Hale's pioneering measurement of sunspot magnetic fields via the Zeeman effect in 1908 (Hale 1908), over a century of progress has been made—from Babcock's photoelectric magnetograph in 1953 (Babcock 1953) to Severny's first vector magnetic field measurement in 1962 (Severny 1962), and subsequently to video vector magnetographs proposed by Beckers and Ramsey (Beckers 1968; Ramsey et al. 1979). Despite these advances, two fundamental bottlenecks persist: (1) model dependence in magnetic field inversion, and (2) limited accuracy in transverse field measurements (Evans 1949; Wang 2003).

AIMS (An Infrared System for the Accurate Measurement of Solar Magnetic Field) is a national major scientific instrument project funded by the National Natural Science Foundation of China, which passed its project acceptance review in October 2025 (Deng et al. 2025). It exploits the fact that Zeeman splitting scales as $\lambda^2$ while spectral line width scales as $\lambda$, making mid-infrared (8–14 $\mu$m) observations ideal for directly resolving Zeeman components without reliance on radiative transfer models. This approach promises to improve transverse field measurement precision by an order of magnitude (Evans 1949).

As a large, complex ground-based telescope (Figure 1), AIMS presents unique challenges for its Master Control System (MCS):

- *Remote unattended site*: The telescope is sited at Saishiteng Mountain, Lenghu, Qinghai Province—a high-altitude plateau (>4000 m) with excellent mid-infrared atmospheric windows but harsh conditions unsuitable for permanent staffing (Bao et al. 2023).
- *Heterogeneous multi-institute integration*: Components are developed by multiple institutes (e.g., Xi'an Institute of Optics and Precision Mechanics for drive systems, Shanghai Institute of Technical Physics for the FTIR spectrometer), each with distinct hardware interfaces (serial, Ethernet, custom protocols).
- *Complex observation modes*: Scientific observations include FTIR fixed-point and scanning modes, polarimetric and non-polarimetric configurations for longitudinal and vector magnetic field measurements, and 8–10 $\mu$m broadband imaging, plus multiple calibration modes.

Existing telescope control frameworks—including INDI, RTS2, ASCOM (George & Denny 2001; Kuba´nek et al. 2008), and observatory-specific systems such as the Gemini OCS (Gillies & Walker 1996), the ASTE remote control system (Ezawa et al. 2004), the New Solar Telescope control system (Yang et al. 2006), and the LAMOST camera-control system (Tian et al. 2018)—each address particular aspects of telescope automation. Significant progress has also been achieved at the Huairou Solar Observing

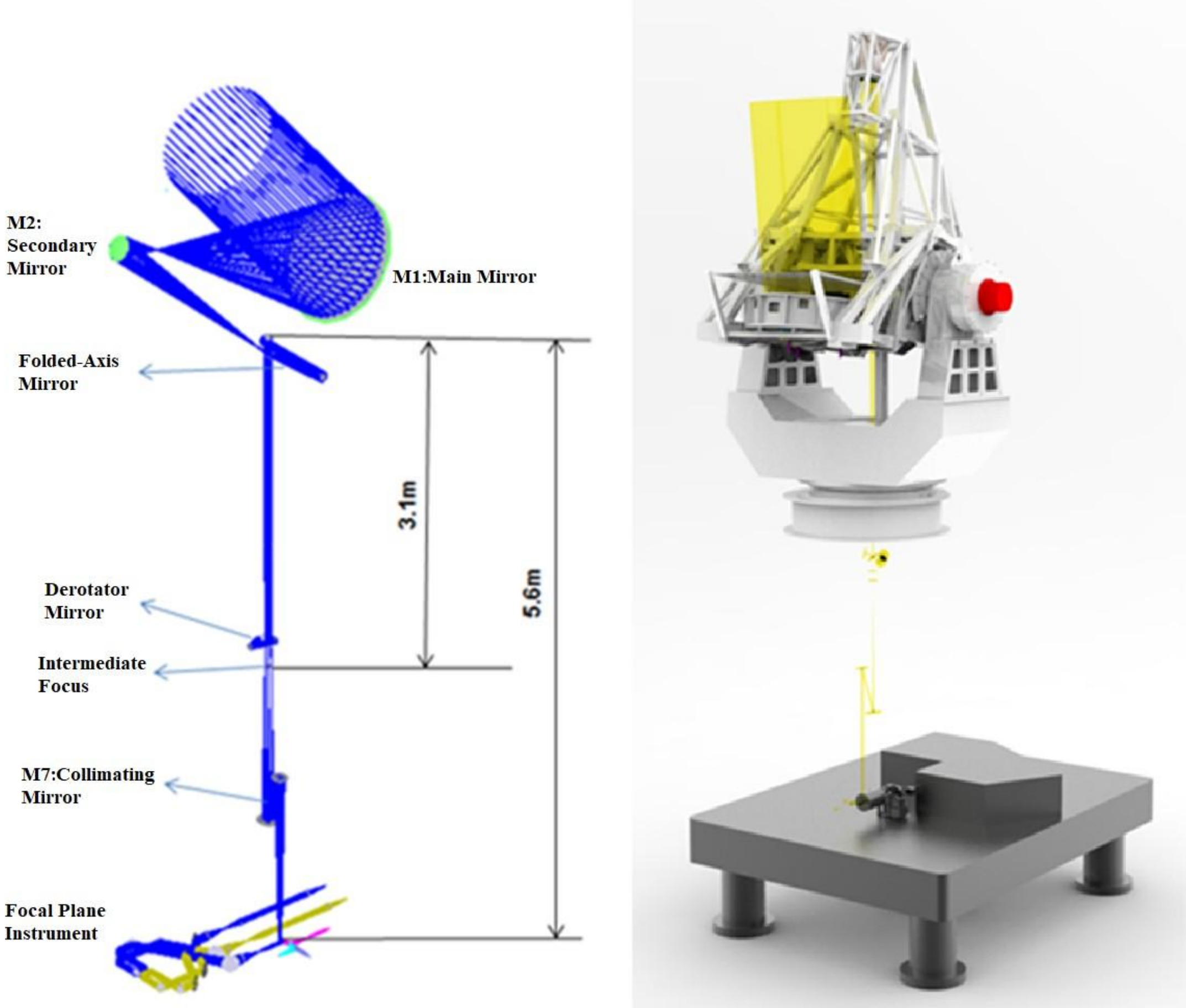

Fig. 1: AIMS optical and mechanical system model.

Station, where automated full-disk magnetic field observations Lin et al. (2013), real-time solar image cross-correlation correction Shen et al. (2013), and full-disk guiding applications have been realized Guo et al. (2016). Nevertheless, these efforts remain focused on localized functional linkage rather than establishing a full-chain, coordinated observation control system that integrates telescope mounts, optical components, detectors, and environmental factors. Consequently, existing solutions cannot fully accommodate AIMS's requirements for heterogeneous hardware integration, progressive autonomous evolution, and complex mid-infrared observation workflows.

In this paper, we present the complete design and implementation of the AIMS MCS. Section 2 describes the requirements analysis and system modeling. Section 3 presents the architecture design, including a novel automation classification, the three-layer system framework, the microservice software architecture, and control/safety strategies. Section 4 details key enabling technologies. Section 5 reports validation results across three telescopes. Section 6 summarizes our contributions and outlines future work.

## 2 REQUIREMENTS ANALYSIS AND SYSTEM MODELING

### 2.1 Autonomy Requirements

The AIMS MCS must address four categories of requirements driven by the telescope's operational context:

*(1) High-altitude remote site constraints* (Figure 2). The mid-infrared atmospheric window at 8–14 $\mu$m is strongly affected by water vapor absorption and atmospheric turbulence. The Saishiteng Mountain site offers low precipitable water vapor and excellent seeing, but its remoteness necessitates autonomous operation with minimal on-site personnel.

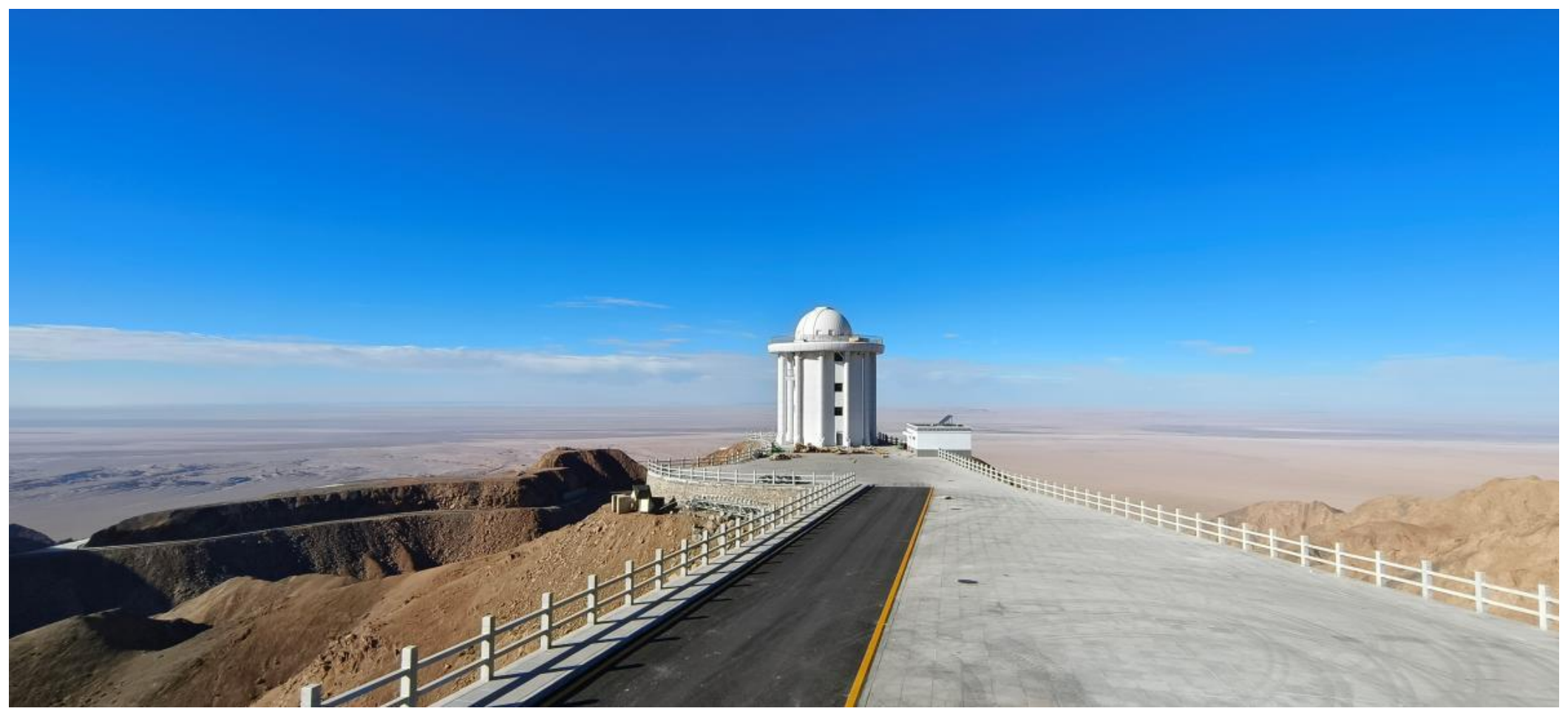

Fig. 2: AIMS observation station located at high altitude.

*(2) Heterogeneous component integration.* AIMS components span multiple hardware interface types—serial (RS-232/485), Ethernet (TCP/IP), CAN bus, and proprietary protocols. The MCS employs a middleware and adapter design pattern: standardized abstract interfaces encapsulate device-specific drivers, enabling the upper control layers to interact through a unified API regardless of the underlying hardware.

*(3) Complex observation modes.* Scientific observations encompass: FTIR observations in six modes—fixed-point non-polarimetric, fixed-point polarimetric, gapped field-scanning non-polarimetric, seamless field-scanning non-polarimetric, gapped field-scanning polarimetric, and seamless field-scanning polarimetric—where the scanning variants arise from whether the scan step size equals (seamless) or exceeds (gapped) the slit width, 8–10 $\mu$m broadband imaging, and calibration modes including flat-field, dark-field, instrument/atmospheric background, and geometric/resolution calibration between the FTIR and broadband camera.

*(4) Multi-user data requirements.* Scientists require environmental and instrument metadata for data calibration; engineers need real-time device health monitoring for performance optimization; operators need rapid situational awareness for intervention decisions.

### 2.2 System Modeling

We model the AIMS controlled components across five spatial zones (Figure 3): (1) the dome (Gregorian primary/secondary mirrors, guider, polarization calibration unit 1, wavefront sensor, coude´ system), (2) the vertical light guide (five-mirror field de-rotation system, polarization calibration unit 2, tip-tilt mirror), (3) the optical platform (image stabilization, guiding/tracking, wavefront correction via hexapod, 8–10 $\mu$m camera, polarization analyzer, FTIR), (4) the control room (observation control, data storage/transfer, database), and (5) outdoor facilities (GPS/BeiDou timing, meteorological/safety monitoring).

The observation workflow (Figure 4) proceeds through: system startup → environmental assessment (external weather and internal equipment checks, repeated every 30 min during daytime) → hardware commissioning (dome, telescope pointing/tracking, thermal control, optical instruments) → software commis-sioning (trial observation and data validation) → science/calibration observation → system shutdown (triggered by sunset, adverse conditions, or critical faults).

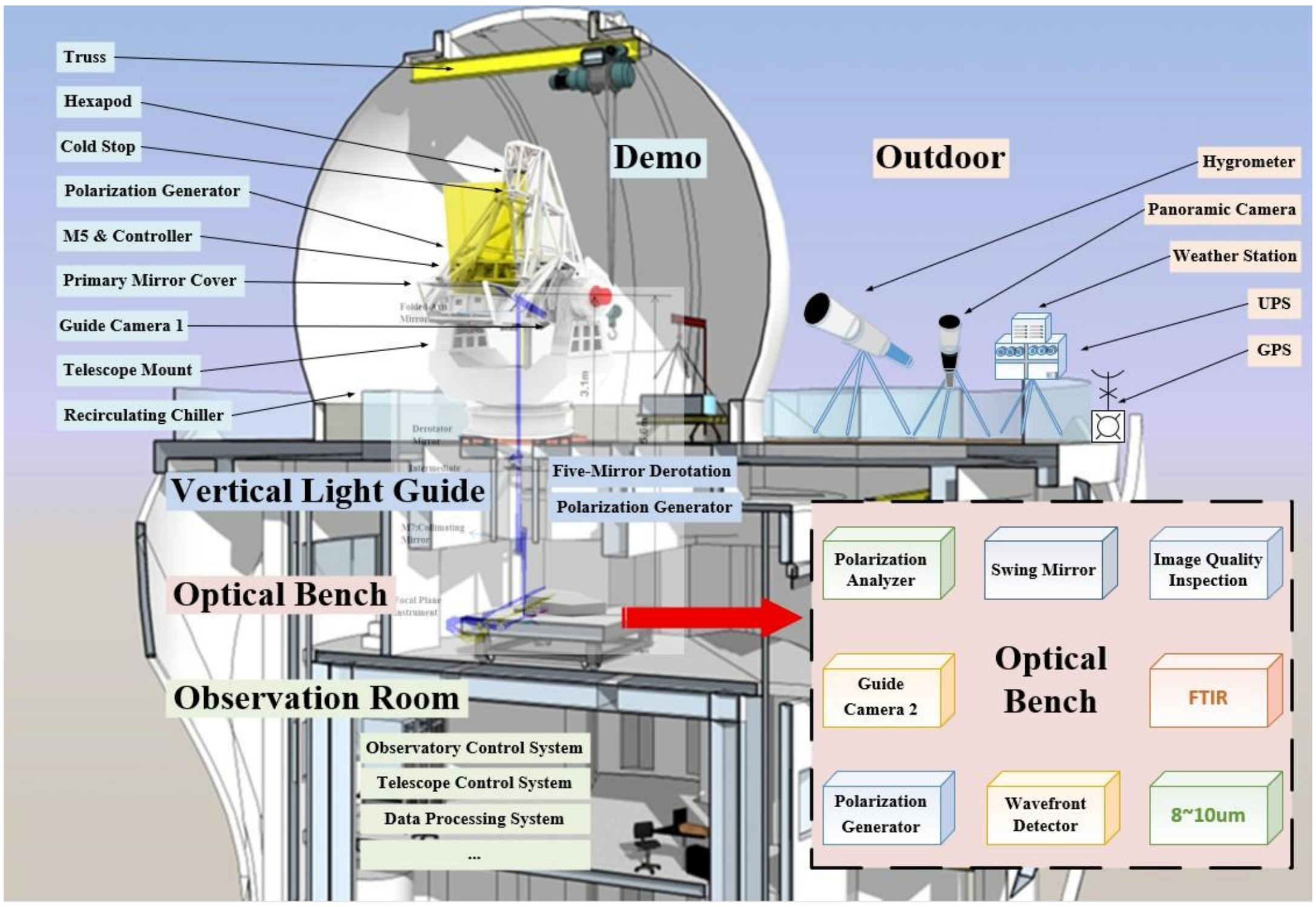

Fig. 3: Spatial model of AIMS controlled components across five zones.

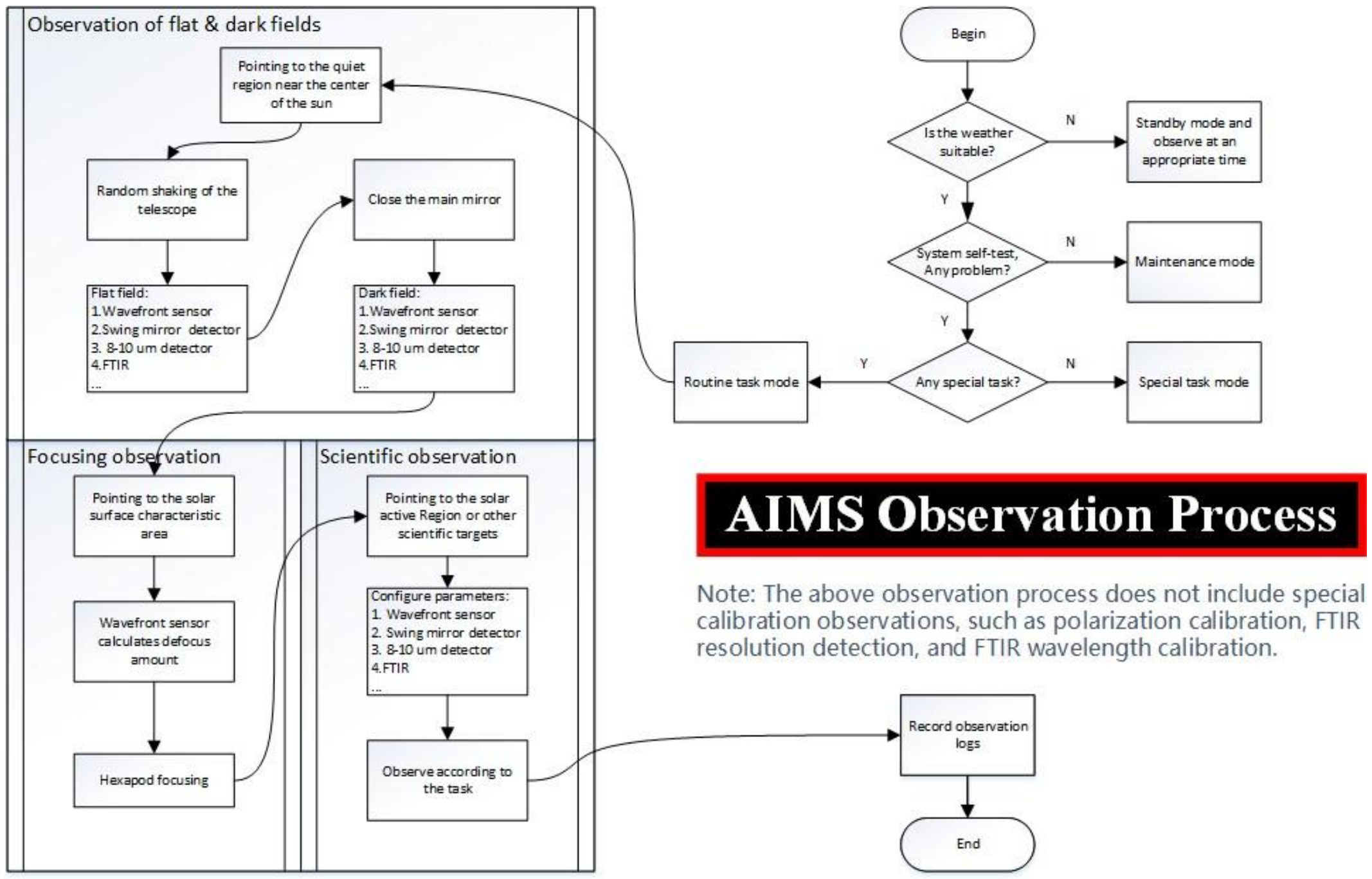

Fig. 4: AIMS observation workflow.

## 3 ARCHITECTURE DESIGN

### 3.1 Telescope Automation Classification (L0–L5)

Existing telescope automation classifications (Castro-Tirado 2010) typically distinguish four levels (automated, remote, robotic, intelligent), but suffer from ambiguous boundaries, lack of upgrade paths, and poor cross-system compatibility. For instance, under the conventional “automated” category, a telescope capable only of auto-exposure and a telescope executing fully automated startup–observation–shutdown sequences are classified at the same level, despite vastly different autonomy capabilities. This ambiguity hinders both system design and cross-facility comparison. Inspired by the SAE J3016 autonomous driving standard, we propose a six-level classification with well-defined boundaries and progressive evolution capability:

- **L0 (Manual)**: All operations performed manually; basic electromechanical actuation only.
- **L1 (Basic Automation)**: Automated pointing, tracking, and simple repetitive tasks; observation planning and complex adjustments require human operation.
- **L2 (Partial Automation)**: Pre-programmed observation sequences with automatic parameter adjustment within limited environmental variations; operator monitoring required.
- **L3 (Conditional Autonomy)**: Autonomous observation under specific conditions (good seeing, known target lists), including automatic scheduling, environmental adaptation, and basic data quality checks; safe shutdown when conditions exceed the operational design domain.
- **L4 (High Autonomy)**: Autonomous operation across a broad operational domain, including real-time strategy optimization, complex data stream processing, and preliminary analysis; operator intervention only for faults or exceptional situations.
- **L5 (Full Intelligent Autonomy)**: Fully autonomous under all foreseeable conditions, incorporating machine learning for strategy optimization, unknown target discovery, self-maintenance, and deep data analysis.

Each level strictly subsumes the capabilities of lower levels, and the framework is designed so that a system architected for L5 can operate at any lower level during phased development. AIMS currently operates at L1, with a roadmap targeting progressive evolution to L5.

Table 1 provides measurable indicators for assessing each automation level, enabling objective cross-system comparison and guiding progressive upgrade decisions.

Table 1: Measurable Criteria for Telescope Automation Levels

| Level | Human Intervention | Env. Awareness | Decision Scope | Fault Recovery |
|---|---|---|---|---|
| L0 | Continuous | None | None | Manual |
| L1 | Frequent (planning) | None | Single tasks | Manual |
| L2 | Periodic monitoring | Limited | Pre-programmed seq. | Basic retry |
| L3 | Exception-only | Active | Condition-based | Safe shutdown |
| L4 | Rare (critical faults) | Predictive | Strategy optimization | Self-recovery |
| L5 | None | Full + learning | AI-driven discovery | Self-maintenance |

### 3.2 Three-Layer System Framework

Reverse-engineering from the target L5 capability and combining the requirements analysis, we design a three-layer framework (Figure 5):

*Device Control Layer* comprises three subsystems that directly interface with hardware: the Telescope Control System (TCS, for pointing, tracking, guiding, and field de-rotation via the five-mirror system), the Instrument Control System (ICS, for focal-plane instruments including the FTIR, polarization analyzer, and cameras), and the Observation Control System (OCS, for data acquisition, preprocessing, and storage).

*Autonomy Support Layer* provides the decision-making context through five subsystems: the Communication Control System (CCS, inter-subsystem messaging and relay), the Power Monitoring System (PMS, energy supply monitoring and management), the Environment Perception System (EPS, meteorological and atmospheric condition monitoring), the Observation Interaction System (OIS, user task scheduling and visualization), and the Model Evolution System (MES, machine-learning-based continuous optimization).

*Central Decision Layer* consists of the Central Decision System (CDS), which integrates information from all subsystems, performs intelligent scheduling and risk assessment, and issues coordinated control commands to the device control layer.

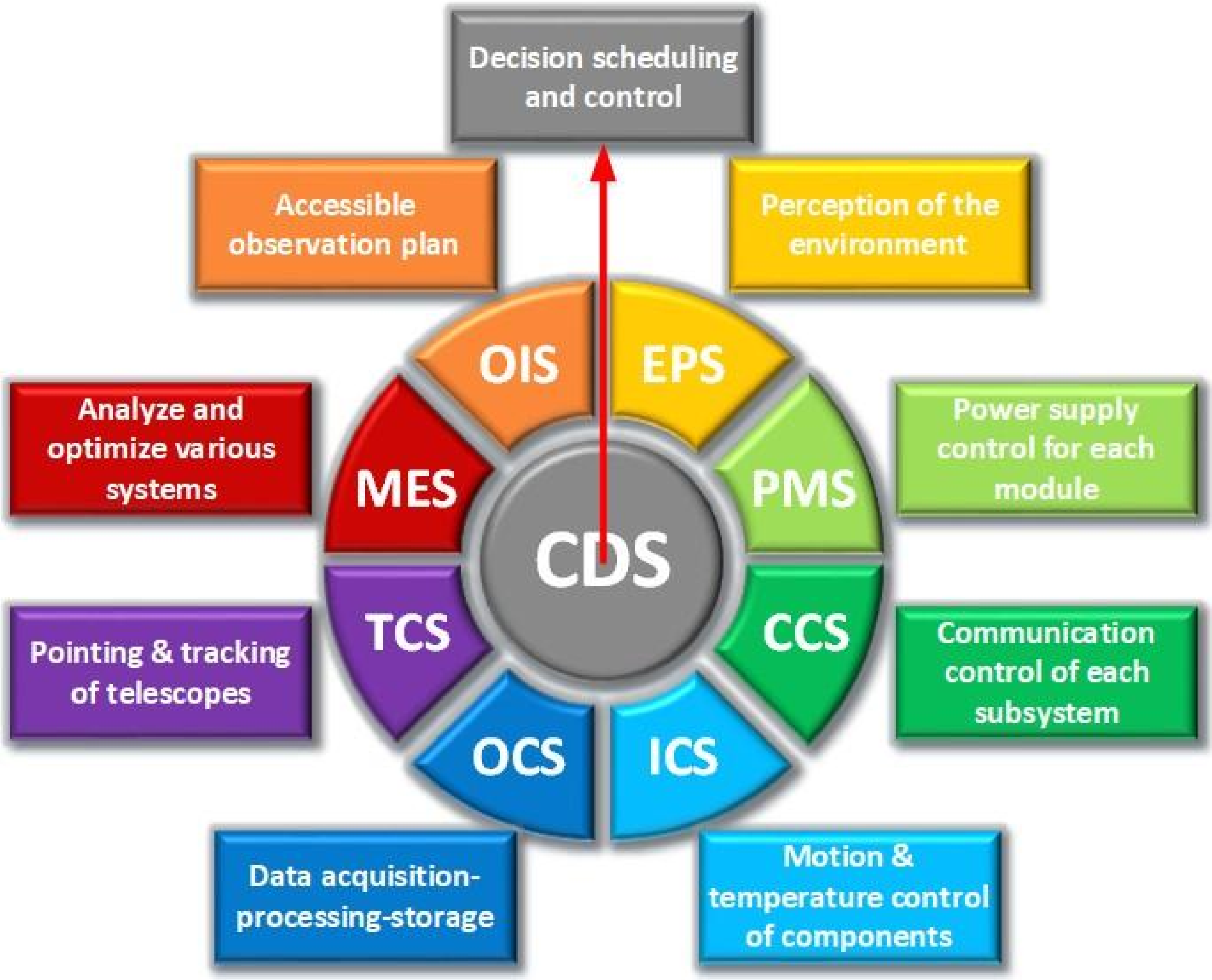

Fig. 5: AIMS MCS system framework overview, showing the three-layer architecture and L0–L5 progressive evolution roadmap.

Each subsystem follows a vertical architecture (Figure 6): hardware → hardware driver → low-level control (function-to-command mapping) → intermediate control (control logic, safety, communication) → top-level control → communication protocol layer → application layer (GUI/console). Subsystems can operate independently for debugging or be coordinated through the CDS for autonomous observation.

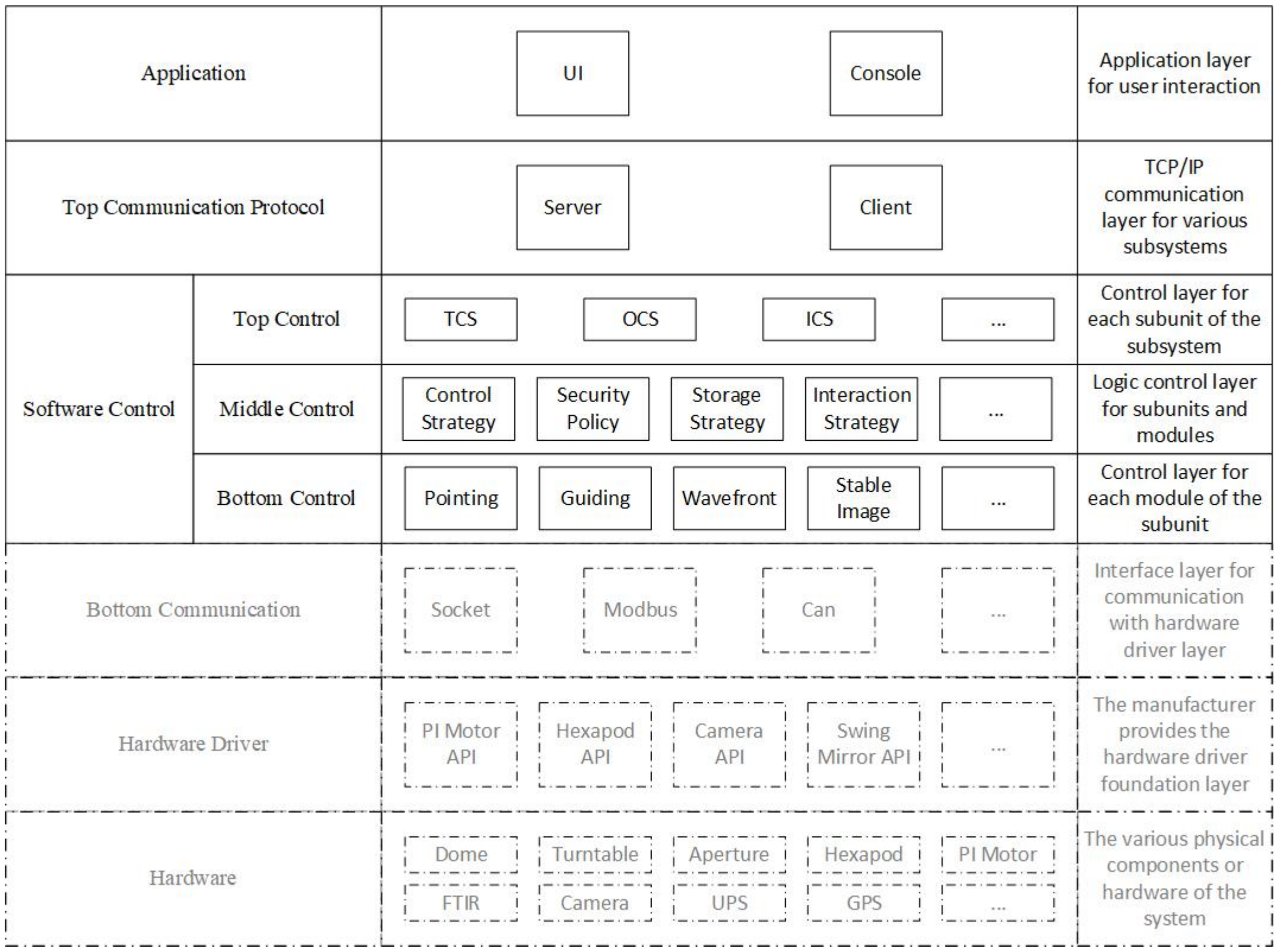

Fig. 6: Vertical architecture of subsystems.

### 3.3 Microservice Software Architecture

Given AIMS's requirements for loose coupling, high cohesion, heterogeneous technology stacks, and continuous integration of components from multiple institutes, we adopt a microservice architecture (Newman 2015). Alternative architectures were evaluated: monolithic (insufficient modularity), layered/SOA (insufficient hardware flexibility), event-driven/serverless (incompatible with real-time hardware control), and containerized/microkernel (prohibitive development cost). Microservices offer process isolation, independent deployment, heterogeneous technology stacks, and flexible delivery cycles—precisely matching AIMS's needs.

The microservice design follows a bottom-up abstraction: individual devices → functional units → groups → subsystems. Each subsystem is an independent service with its own database, state monitoring, fault handling, and UI, communicating via the lightweight TCP/IP protocol defined by the CCS.

The CCS protocol employs a structured message format comprising ten mandatory fields: protocol version, message type, send/receive category, auxiliary attributes, message length, sender identifier, receiver identifier, priority level, message content, and SHA-512 integrity checksum. Single messages are limited to 256 KB; larger payloads are automatically segmented and reassembled. For file transfers, metadata (file-name, size, unique ID, checksum) is sent prior to the file content, eliminating per-block verification overhead. Communication latency over the local-area Ethernet network is <100 ms, well within the requirements of hardware control cycles (serial devices: 50–500 ms response time). The protocol supports RSA key exchange and AES encryption at the code level; in current deployment on isolated LANs, encryption is disabled to minimize computational overhead on real-time control loops, while Role-Based Access Control (RBAC) remains active for function-level authorization. Each subsystem client logs instrument state at 1-second intervals to local storage, providing comprehensive fault traceability for remote diagnosis—critical for the high-altitude AIMS site where on-site access is limited.

### 3.4 Control Strategies and Safety

*Multi-dimensional control.* Observation task execution follows a priority hierarchy: energy status (highest priority) → environmental conditions → observation task scheduling → fault handling. The CDS evaluates candidate tasks from the OIS queue based on priority, estimated duration, energy consumption, environmental sensitivity, and system health, then dispatches commands to the device control layer.

*Communication and security.* Inter-subsystem communication uses a lightweight custom TCP/IP protocol with structured messages (header, body, tail) and integrity verification. Communication security combines asymmetric key exchange with symmetric encryption, while Role-Based Access Control (RBAC) restricts system functions to authorized user roles. Comprehensive state and execution logging enables fault traceability and rapid diagnosis.

*Safety and fault handling.* The system implements a priority-based fault response hierarchy: energy anomaly (highest priority) → environmental hazard → equipment fault → observation anomaly (lowest). For AIMS, the thermal stop temperature is continuously monitored against the ambient environment temperature; if the thermal stop exceeds the ambient by more than a seasonally-adjusted threshold (e.g., 5°C), the system immediately closes the primary mirror protection curtain and thermal stop cover to prevent optical damage. On SFMM, when the PMS detects a transition from mains power to UPS/battery supply, the CDS evaluates remaining battery capacity against the estimated safe-shutdown duration; if insufficient, autonomous shutdown is initiated immediately, with automatic restart upon mains power recovery contingent on favorable weather conditions. The microservice architecture inherently provides fault isolation: a failure in any single subsystem is contained within that service's process boundary, allowing other subsystems to continue operating while the CDS reconfigures the observation plan.

## 4 KEY ENABLING TECHNOLOGIES

Achieving autonomous telescope operation requires not only the architectural innovations described in Section 3, but also the automation of several critical observation functions that traditionally depend on experienced operators. In the AIMS context, these include: precise solar acquisition without manual intervention, maintaining optimal focus under turbulent atmospheric conditions, and continuous adaptation to changing environmental conditions such as cloud cover, power supply fluctuations, and thermal variations in the optical path. The following subsections detail our solutions to each of these challenges.

### 4.1 Automatic Pointing and Tracking

Precise automatic pointing and tracking form the foundation of autonomous observation. The pointing system computes the solar position using the VSOP87 planetary theory (Liu 2011), converting from heliocentric ecliptic coordinates to local horizontal coordinates (azimuth $A$, elevation $H$) via:

$$\sin H = \sin\phi \sin\delta + \cos\phi \cos\delta \cos t, \tag{1}$$

where $\phi$ is the observer's latitude, $\delta$ the solar declination, and $t$ the hour angle.

Instrumental axis errors arising from manufacturing tolerances, thermal deformation, and structural flexure are corrected using a composite model combining spherical harmonic fitting (Wang et al. 2012) with a fixed sky-region model. The spherical harmonic expansion:

$$F(\theta, \lambda) = \sum_{l=0} \left( A_l^0 P_l^0(\cos\theta) + \sum_{m=1}^{l} [A_l^m \cos(m\lambda) + B_l^m \sin(m\lambda)]\, P_l^m(\cos\theta) \right) \tag{2}$$

captures the spatial distribution of pointing errors across the sky. This model has been validated on the WenQuan telescope (pointing RMS < 20″) and the Solar Full-disk Multi-layer Magnetograph (SFMM) (pointing RMS < 5″) (Tong et al. 2024).

Closed-loop guiding uses a solar full-disk image from the guide camera. The system computes the solar disk centroid displacement between consecutive frames and applies PID-controlled corrections to the telescope drive, achieving guiding RMS < 1″ over 30 min (Figure 7). An adaptive auto-exposure mechanism (Section 4.3) maintains guide image quality during thin-cloud passages.

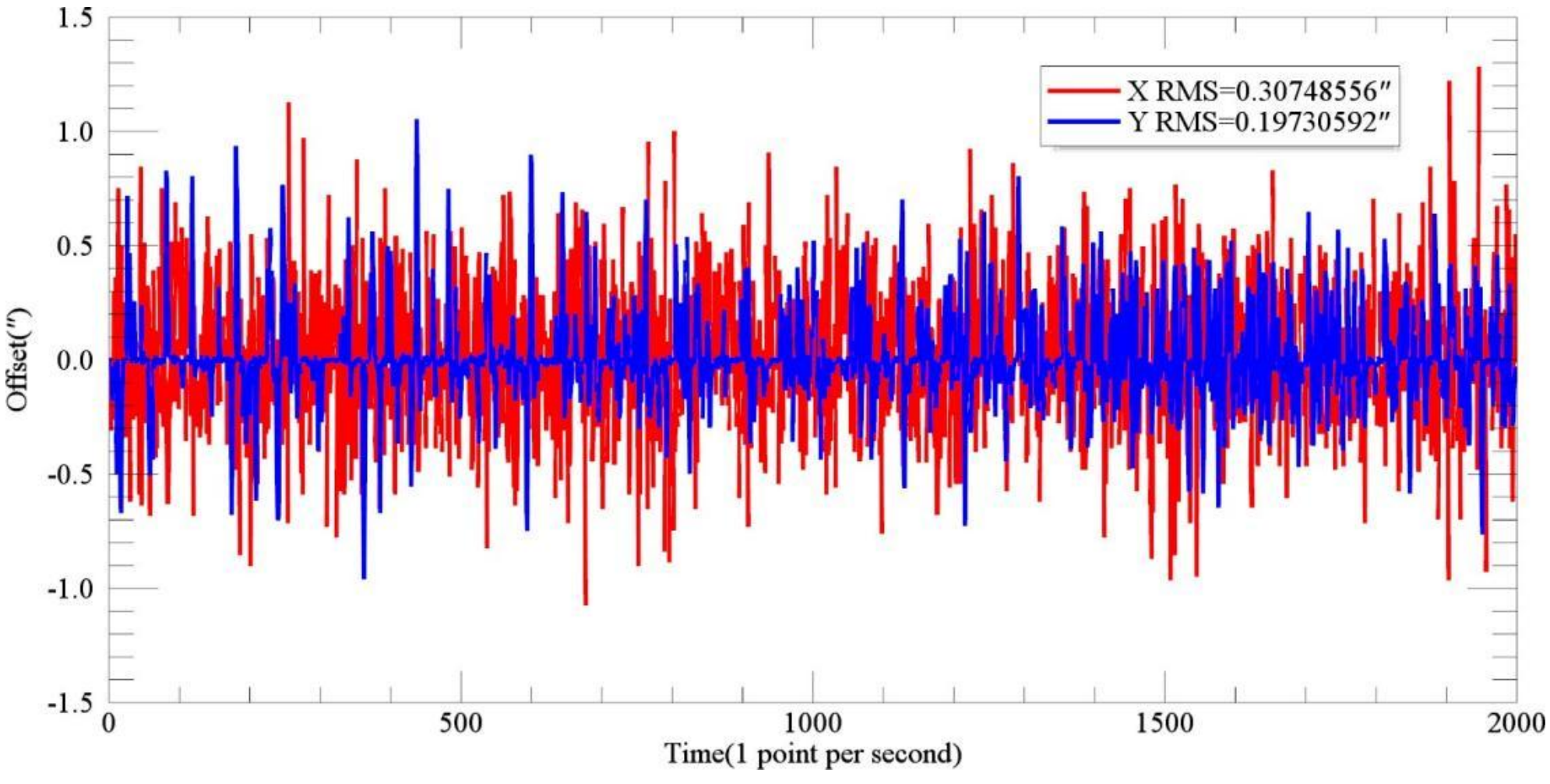

Fig. 7: Optimized guiding results on SFMM, demonstrating RMS < 1″/30 min.

### 4.2 Automatic Focusing

Solar autofocus faces unique challenges: atmospheric turbulence introduces strong temporal variability and multi-scale spatial variations, while instrument errors (tracking jitter, filter bandpass drift, optical non-uniformities) further complicate image quality assessment. Conventional gradient-based, entropy-based, or simple spectral methods fail under these conditions.

We developed a method combining three techniques:

*(1) Lucky-frame selection.* At each focus position, multiple short-exposure frames (exposure time < atmospheric freeze time, typically <10 ms) are acquired. The best-quality frame is selected using the power spectral ratio method, suppressing temporal turbulence effects.

*(2) Power spectral ratio analysis.* Each selected frame is zero-mean normalized, Fourier-transformed to obtain the power spectrum $P(u, v)$, converted to polar coordinates, and azimuthally averaged to yield $L(\rho)$. The spectral ratio $C_n(\rho) = L_n(\rho)/L_0(\rho)$ relative to a reference frame is evaluated at the telescope's spatial resolution limit (Liang et al. 2014) (Figure 8).

*(3) Gaussian fitting.* The spectral ratio values across focus positions are fitted with a Gaussian function $G(x) = a\exp[-(x-b)^2/(2c^2)]$, and the optimal focus position is determined from the fitted peak $b$.

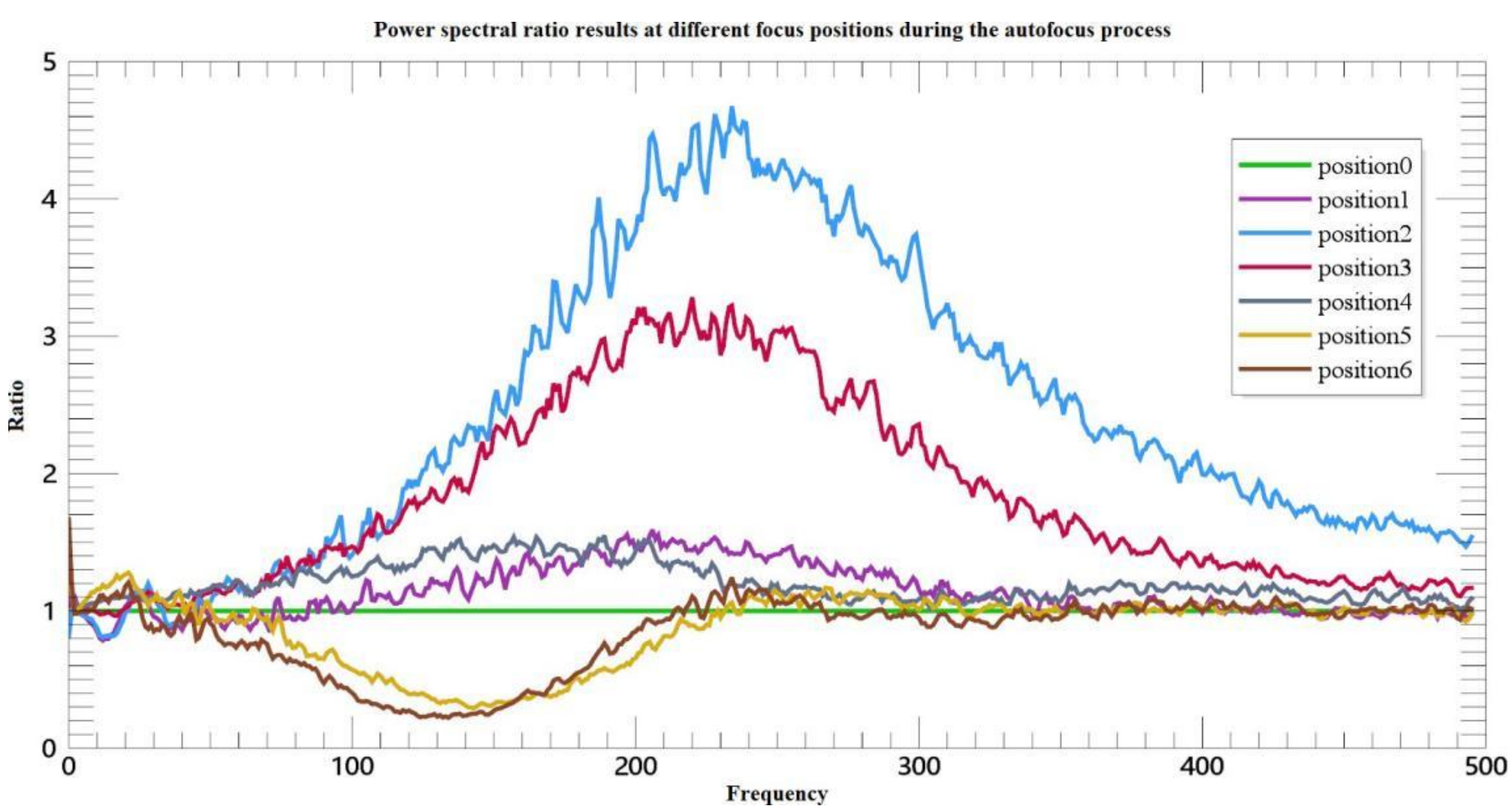

Fig. 8: Power spectral ratio results at different focus positions during the autofocus process.

This method has been validated on both the WenQuan telescope and SFMM, achieving reliable autofocus under varying atmospheric conditions.

## 4.3 Environment-Adaptive Observation

Autonomous observation requires the MCS to continuously sense and respond to environmental changes—from transient cloud passages and solar irradiance fluctuations to power supply anomalies and thermal variations in the optical path. We integrate three complementary capabilities into a unified environment-adaptive framework.

*(1) Solar-disk-targeted auto-exposure.* Standard auto-exposure algorithms optimize for the entire image frame, which is unsuitable for solar observation where the bright solar disk occupies only a portion of the field. We developed a strategy that applies a brightness threshold to isolate the solar disk region, restricts exposure optimization to that region only, and employs a frame-skipping mechanism to execute adjustments on non-science frames, preventing frame-rate degradation during continuous observation. Configurable brightness target ranges and gain/exposure bounds allow customization for different scientific objectives. This approach is particularly effective during thin-cloud passages, where increased exposure compensates for reduced solar intensity, maintaining guide image quality for closed-loop tracking.

*(2) Cloud detection and weather assessment.* The Environment Perception System (EPS) integrates an all-sky camera for cloud distribution monitoring, a pyranometer for solar irradiance measurement, and cloud recognition algorithms (Yang et al. 2012; Urquhart et al. 2015; Wang et al. 2023) for real-time cloud coverage analysis. The combined data enable the CDS to determine whether current conditions permit observation, predict near-term cloud evolution for observation scheduling, and record cloud events as metadata for post-observation data quality assessment (Figure 9).

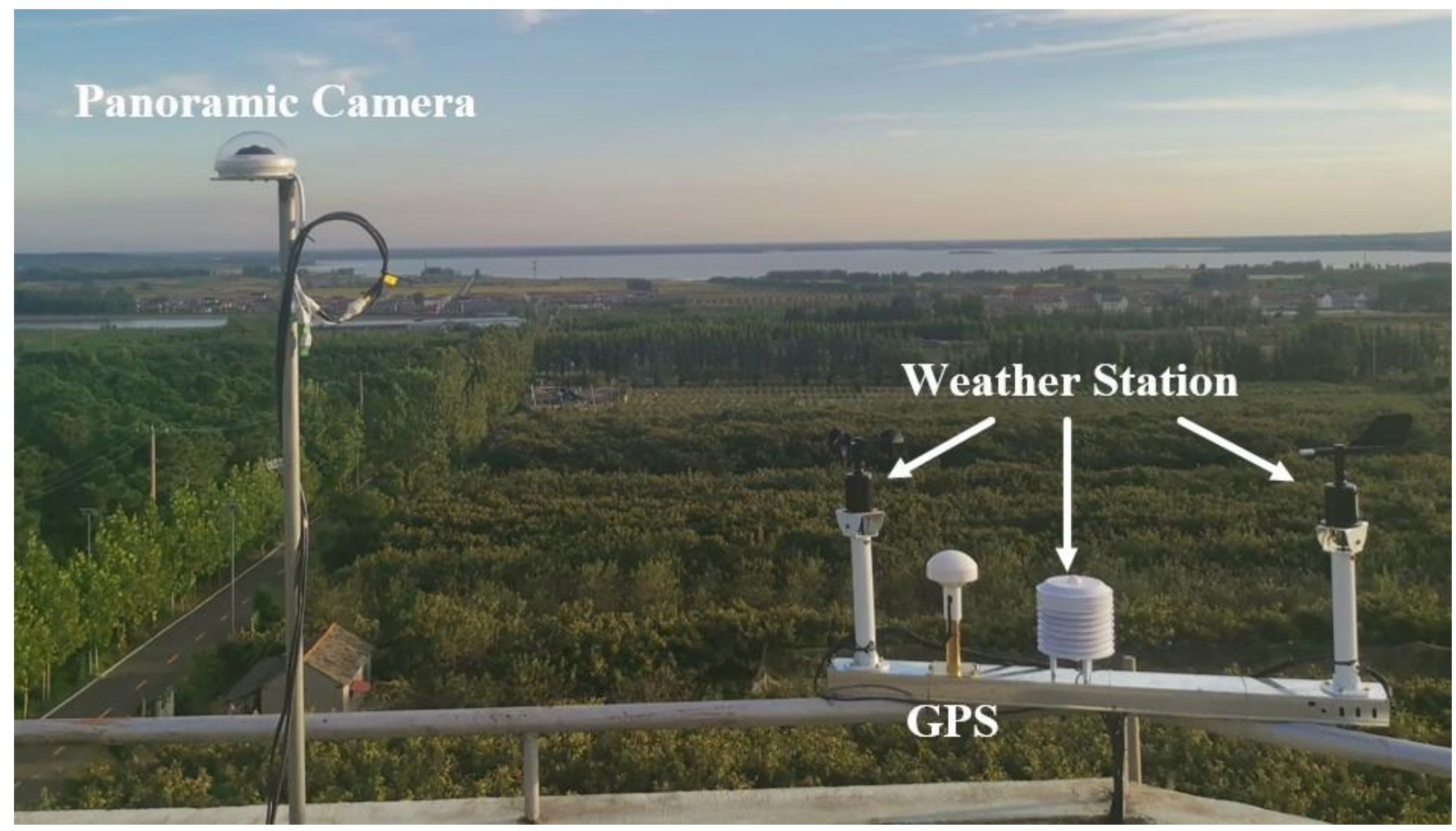

Fig. 9: Deploying weather station and panoramic cloud camera in SFMM.

*(3) Power and thermal monitoring.* The Power Monitoring System (PMS) provides real-time monitoring of the facility's energy supply, including UPS and battery status, mains power quality, and per-subsystem power consumption. On SFMM, the PMS feeds directly into the CDS state machine: when power anomalies are detected, the system autonomously transitions to a safe shutdown state and resumes observation upon power recovery. For AIMS, the coude´ room houses precision optical components (polarization analyzer, FTIR) that require constant temperature and humidity. A dedicated thermal monitoring program continuously tracks the air conditioning system's output, triggering alerts or observation suspension if environmental conditions drift outside the specified tolerance, thereby protecting both optical alignment stability and data quality.

## 5 APPLICATIONS AND VERIFICATION

The AIMS MCS design has been validated across three telescopes at progressive automation levels (Figure 10), providing evidence for the applicability of the framework across different telescope configurations.

Table 2 summarizes the implementation status of each major function across the three telescopes, clearly distinguishing between functions implemented on AIMS, functions validated on WenQuan/SFMM, and functions planned for future AIMS integration.

Table 2: Implementation Status of MCS Functions Across Three Telescopes

| Function | AIMS | WenQuan | SFMM | AIMS Future |
|---|---|---|---|---|
| Microservice architecture & CCS | ✓ | ✓ | ✓ | — |
| Automated polarization calibration | ✓ | — | — | — |
| FTIR observation automation (6 modes) | ✓ | — | — | — |
| Automatic pointing/tracking | ✓ (standalone) | ✓ (in MCS) | ✓ (in MCS) | MCS integration |
| Closed-loop guiding (PID) | ✓ (standalone) | ✓ (in MCS) | ✓ (in MCS) | MCS integration |
| Autofocus (spectral ratio) | — | ✓ | ✓ | Planned |
| Auto-exposure (solar-disk-targeted) | — | ✓ | ✓ | Planned |
| Automated observation sequences | — | ✓ | ✓ | Planned |
| Cloud detection & weather assessment | — | — | ✓ (deployed) | Planned |
| Power monitoring & autonomous shutdown | — | — | ✓ (deployed) | Planned |
| CDS state-machine decision | — | — | ✓ (deployed) | Planned |
| Telescope axis integration into MCS | — | ✓ | ✓ | In progress |

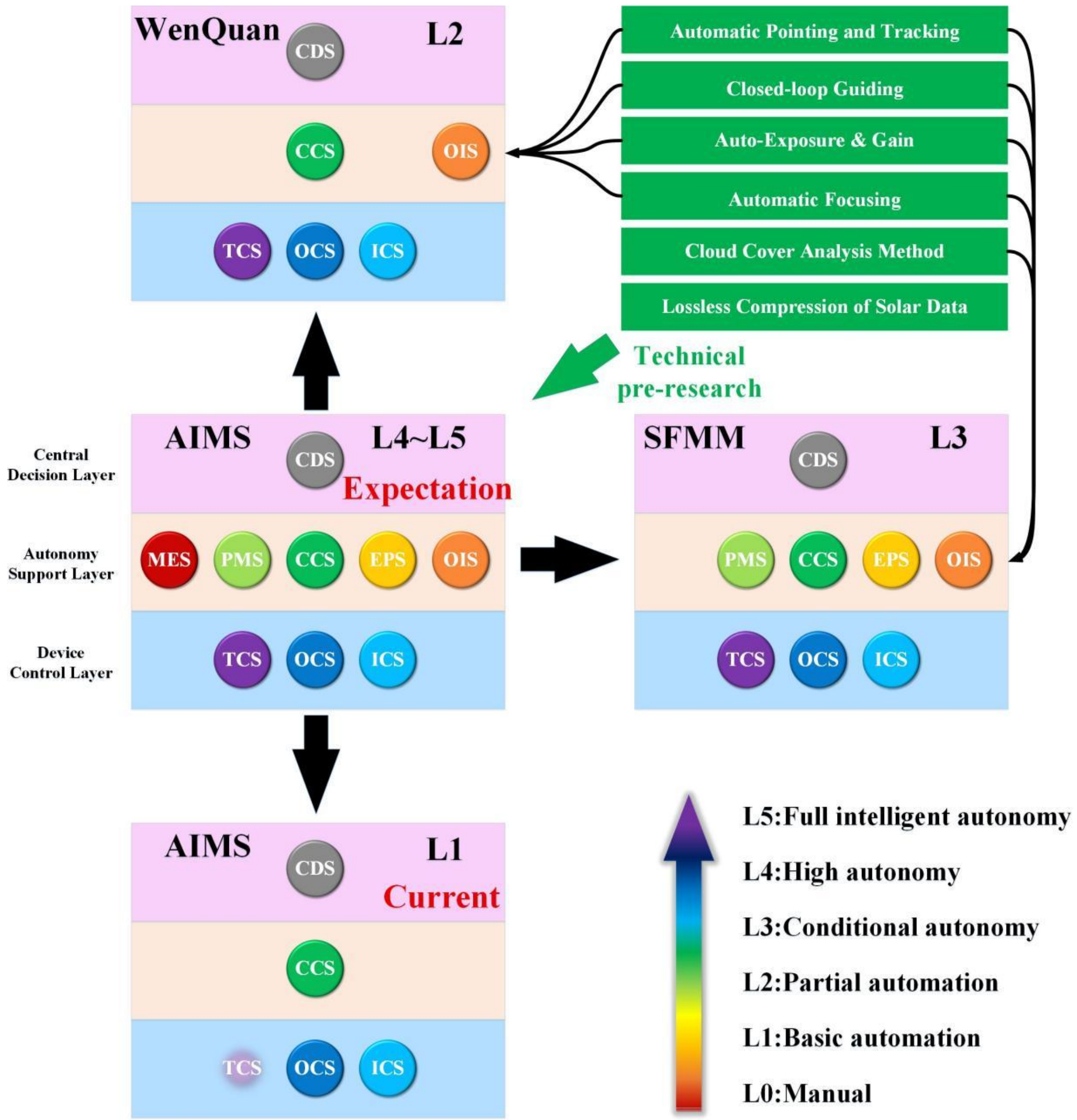

Fig. 10: Application of the AIMS MCS across three telescopes at different automation levels.

### 5.1 AIMS (L1 — Progressive Integration)

AIMS passed its project acceptance review in October 2025. The MCS is progressively integrating subsystem interfaces from multiple collaborating institutes as they become available. Currently integrated components include the dome (primary mirror protection, thermal stop, hexapod), polarization calibration units 1 and 2, the polarization analyzer, the point-source detector, the 8–10 $\mu$m broadband camera, and the FTIR control system. The CCS is operational, while remaining autonomy support subsystems are being integrated as vendor interfaces are delivered.

Within this progressive integration, we have achieved L1 automation through a systematic approach of command mapping, command orchestration, and observation mode combination. However, the legacy manual observation workflow presented significant challenges. The complexity arose not only from the combinations of different polarization modes but also from the stringent coordination required with the FTIR. Due to observation window constraints, the workflow necessitated coordinated horizontal and vertical field scanning using the M11 mirror. Furthermore, deep integration stacking settings at identical acquisition positions were required to ensure signal-to-noise ratios. These factors resulted in a workflow characterized by numerous branches and significant time consumption, where manual operation carried a high risk of error, potentially compromising precious acquisition time. To address this, we adopted the following automation strategies:

*Command mapping and function packages.* We abstract each hardware operation into a callable function and map it to a concise command with well-defined object, content, and completion criteria. These functions, together with their associated state parameters, are encapsulated into *function packages*—the minimum schedulable units of the automation framework. During execution, the system sequentially invokes function packages, verifies completion via state parameters, and proceeds to the next package only upon confirmed success. This mechanism provides high scalability and full traceability of the automation process.

*Automated polarization calibration.* Calibrating the instrumental polarization of the entire telescope requires coordinated operation among multiple distributed units (Hou 2019): the Dome Polarization Calibration Unit, the Vertical Light Guide Polarization Calibration Unit, the Optical Platform Polarization Analyzer, and the Point Source Detector. To manage this multi-unit synergy, we employed a method combining command mapping with serialized encapsulation. Specific motor positioning, data acquisition, and state verification steps are assembled into automated calibration tasks. The system operates independently or as a module within the ICS via TCP connection, ensuring consistent calibration quality without manual intervention.

*Conditional FTIR observation automation.* The six FTIR modes (fixed-point/gapped-scanning/seamless-scanning × polarimetric/non-polarimetric) involve substantial repetitive operations. For example, FTIR fixed-point polarimetric observation requires the field-scanning motor to remain at a preset position while the polarization analyzer cycles through multiple polarization states, with the FTIR acquiring data at each state. By composing function packages into observation sequences with state-parameter verification at each step, we automate these complex multi-device coordination workflows (Figure 11), significantly reducing the required on-site personnel from 7–8 to 2 while also enabling automatic recovery from observation anomalies (Figure 12).

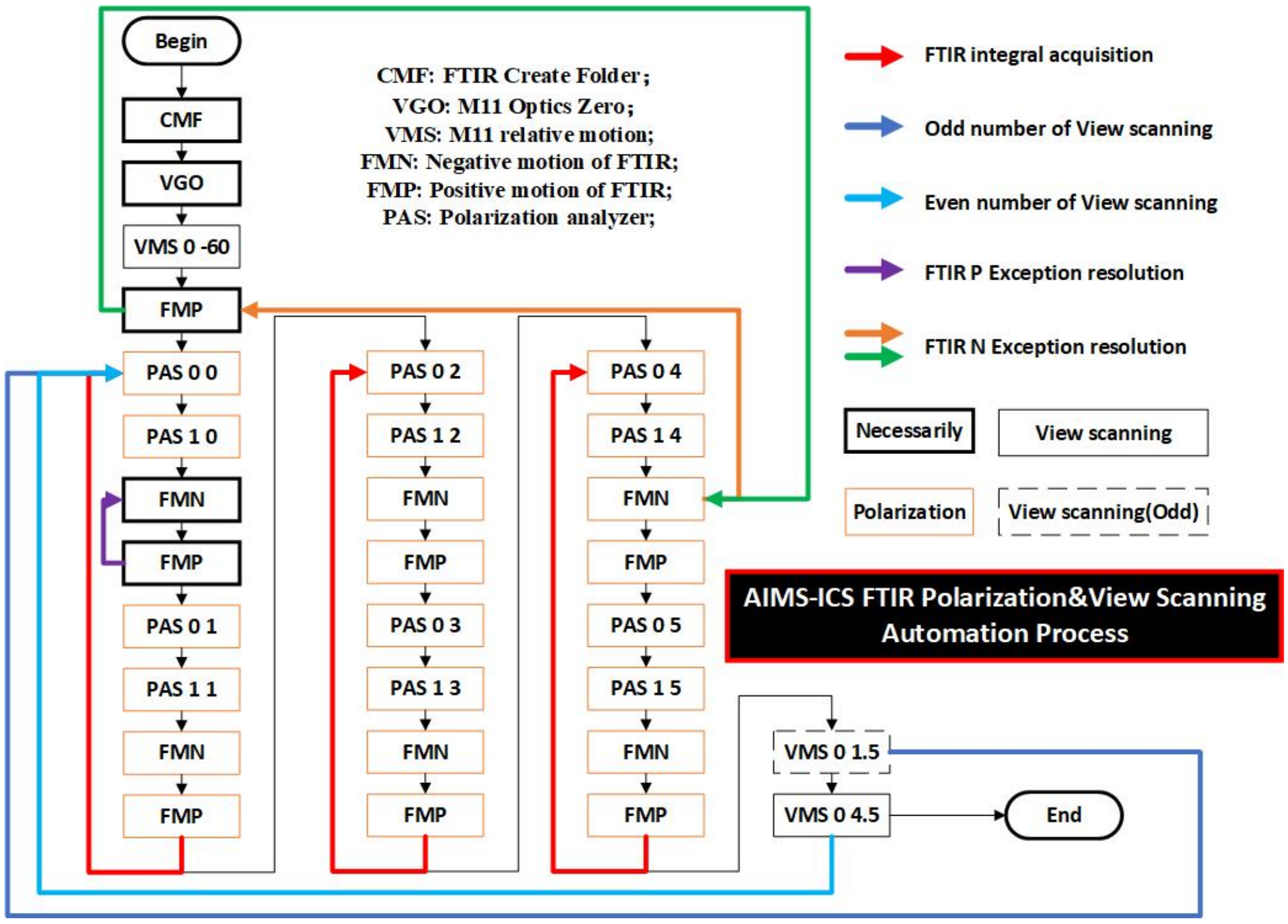

Fig. 11: AIMS scientific and calibration automation process, illustrating command mapping, function package orchestration, and observation mode combination for the six FTIR modes.

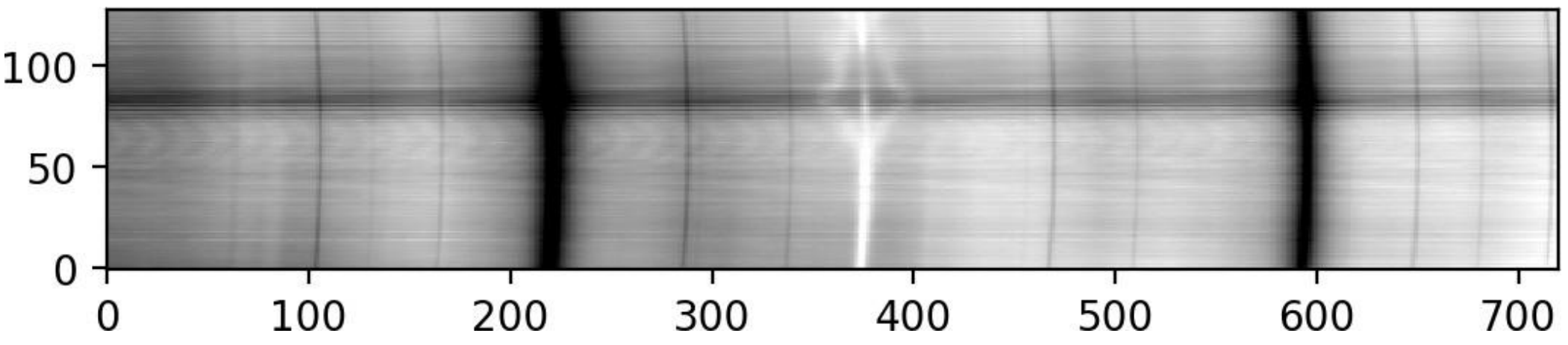

Fig. 12: Processed FTIR observation data.

Table 3: AIMS FTIR Observation Efficiency: Before and After Automation

| Metric | Manual Operation | After Automation |
|---|---|---|
| Required on-site personnel | 7–8 | 2 |
| Data sets per day (field-scanning non-pol.) | 1–2 | 7–8 |
| Time per data set | Several hours | ~ 1 hour |
| Operation error rate | High | Near-zero |

### 5.2 WenQuan Solar Magnetic Field Telescope (L2)

The WenQuan telescope, deployed at Wenquan County, Xinjiang, is China's second solar magnetic field telescope, providing routine full-disk vector magnetograms for space weather forecasting. Following an upgrade in 2022–2023, we implemented a simplified AIMS framework with the same microservice architecture, organized into four subsystems: guiding computation, axis control, focal-plane control, and observation control. The observation workflow (Figure 13) follows a startup–observation–shutdown sequence with automatic pointing/tracking, auto-exposure, and autofocus integrated throughout(Tong 2024).

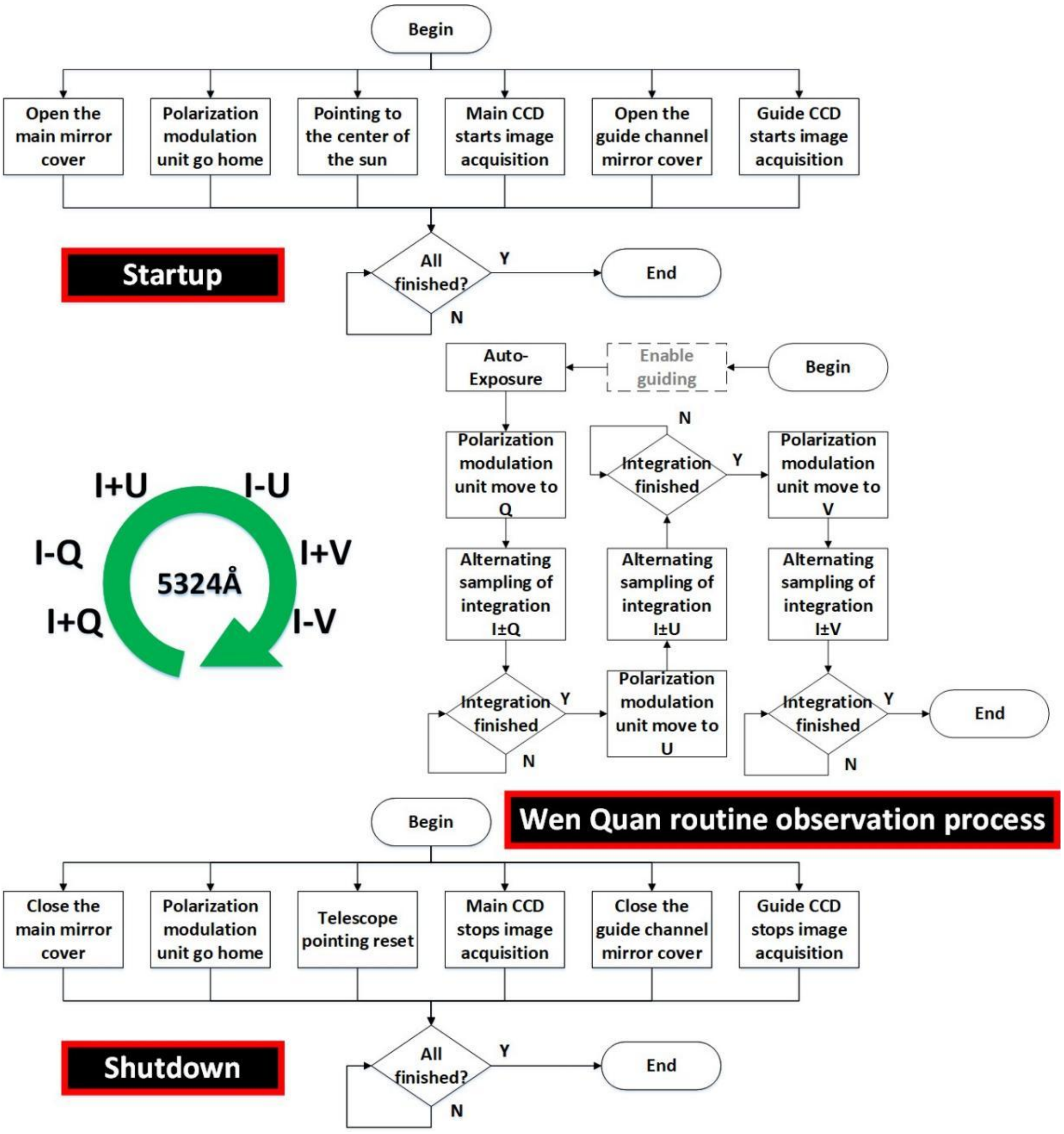

Fig. 13: Observation workflow of the WenQuan solar magnetic field telescope.

The system achieves L2 partial automation: operators need only assess environmental conditions and initiate the system; all observation sequences execute automatically. The system has operated stably for over three years (since February 2023) at the Wenquan Meteorological Bureau, producing reliable data output with operator-friendly interfaces. The autofocus algorithm is executed on a monthly cycle, maintaining optimal image quality across seasonal temperature variations. The pointing and guiding subsystems have remained stable throughout the operational period without requiring recalibration of the axis error model.

### 5.3 SFMM — Solar Full-disk Multi-layer Magnetograph (L2/L3)

SFMM, deployed at Ganyu, Jiangsu Province as part of the Chinese Meridian Project Phase II, performs multi-wavelength observations through four spectral lines (Fe I 5324.19 Å, H$\beta$ 4861.34 Å, H$\alpha$ 6562.8 Å, Ca II 8542.1 Å) across two channels, achieving photospheric vector magnetic field temporal resolution <15 min and chromospheric measurements <180 s. A flare mode provides H$\alpha$/Ca II 8542 Å imaging at 1 fps with 4k×4k resolution.

The SFMM MCS implements the full AIMS framework (excluding MES) with the same microservice architecture, organized into seven subsystems: TCS, ICS, OCS, CCS, PMS, EPS, and CDS. The CDS implements state-machine-based autonomous decision-making (Figure 14): it monitors power supply status and cloud coverage, autonomously deciding whether to observe based on energy reserves and weather conditions. When power anomalies or persistent cloud coverage are detected, the system autonomously executes safe shutdown; when conditions recover, it autonomously restarts observation.

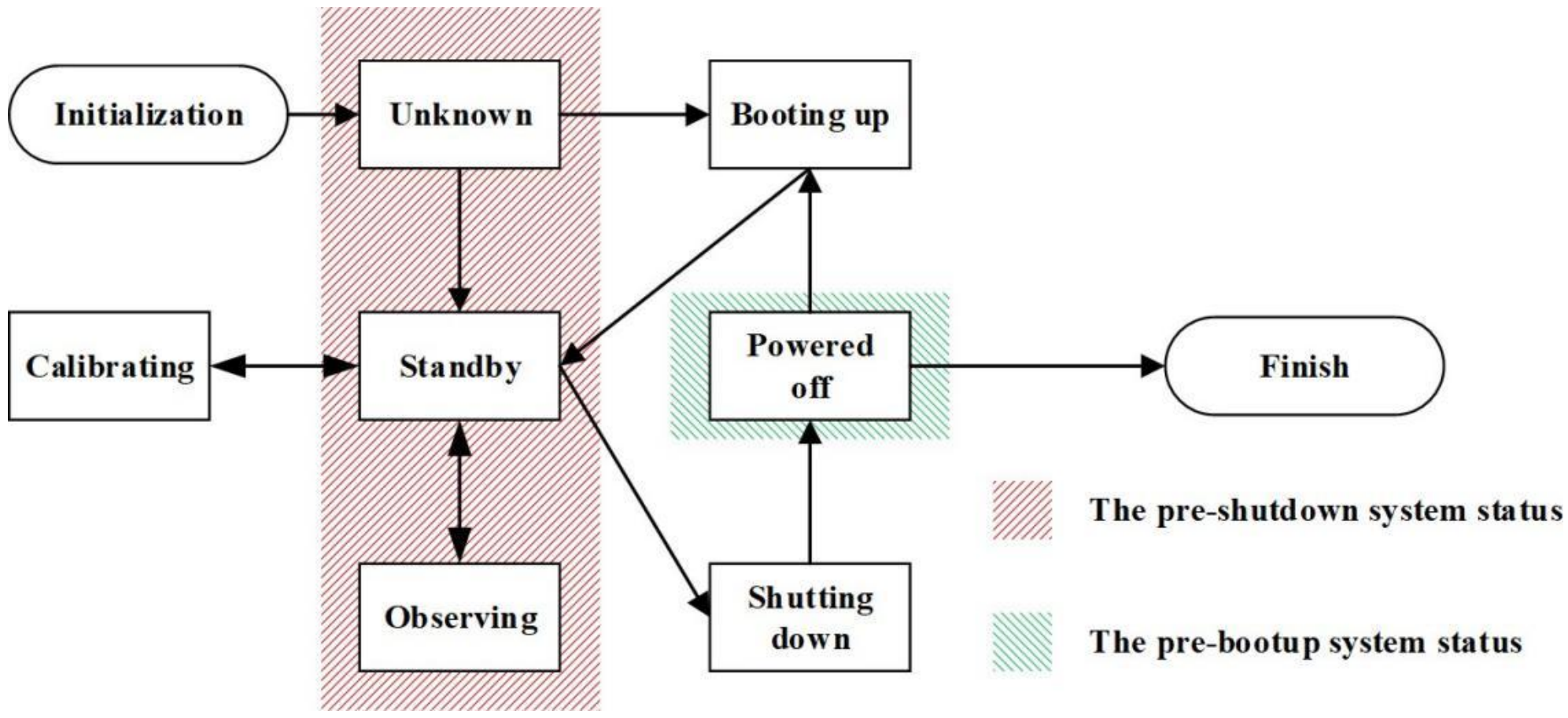

Fig. 14: State transition diagram of the SFMM CDS, showing autonomous startup–observation–shutdown decision flow.

SFMM has achieved stable L2 automation for routine daily operation over 2.5 years since October 2023, producing reliable data output as required by the Chinese Meridian Project operational protocols. Additionally, L3 conditional autonomy functions—including cloud-based observation suspension/resumption, UPS-triggered autonomous shutdown with battery-capacity evaluation, scheduled/periodic task execution, and weather-contingent automatic restart—have been implemented, deployed on-site, and verified through dedicated on-site testing. These functions remain available but are not activated in routine operation due to the project's standardized operational requirements.

### 5.4 Path to High-Level Automation for AIMS

While the current MCS has established a solid foundation for L1 automation, achieving high-level automation and intelligence for AIMS requires further development in system integration and collaborative control.

Future work will focus on integrating remaining subsystem programs, particularly the telescope axis system and dome control systems. A critical challenge lies in breaking down the proprietary software silos of various vendors to establish a unified control interface. This integration will enable coordinated observation among the dome, axis system, environment perception, and energy storage systems. By facilitating real-time data exchange and joint decision-making across these subsystems, the MCS will evolve to support safer, more flexible, and intelligent observation capabilities, ultimately paving the way for the transition from L1 to higher autonomy levels (L3–L5) as outlined in our classification framework.

## 6 SUMMARY AND FUTURE PROSPECTS

We have presented the complete design and implementation of a microservice-architecture Master Control System for the AIMS. The key contributions are:

(1) An L0–L5 telescope automation classification with well-defined level boundaries and progressive evolution capability, inspired by the SAE J3016 autonomous driving standard, providing a practical reference for telescope control system development.

(2) A three-layer system framework (device control, autonomy support, central decision) implemented with a microservice software architecture, achieving loose coupling, high cohesion, and seamless integration of heterogeneous components from multiple institutes.

(3) Solar-observation-specific key technologies: automatic pointing/tracking with composite axis error correction (spherical harmonics + fixed sky-region model), autofocus combining lucky-frame selection with power spectral ratio analysis and Gaussian fitting, and environment-adaptive observation integrating solar-disk-targeted auto-exposure, cloud detection, and power/thermal monitoring for improved operational continuity under variable conditions.

(4) Progressive validation across three telescopes: AIMS (L1, >2 years stable operation), WenQuan (L2, >3 years stable operation), and SFMM (L2 routine operation >2.5 years, with L3 autonomous functions verified), providing practical evidence for the framework's applicability across different telescope scales and automation levels.

Future work targets progressive elevation to L4 automation as remaining subsystem interfaces are integrated, with flexible observation strategies driven by real-time environmental optimization. The ultimate L5 goal envisions the Model Evolution System leveraging multimodal large language models for intelligent observation planning and autonomous scientific discovery. Additionally, construction of a digital twin through multi-band visual fusion is planned to enable comprehensive system simulation and optimization.

**Acknowledgements** We would like to thank the staff at the National Astronomical Observatories, Xi'an Institute of Optics and Precision Mechanics, Shanghai Institute of Technical Physics, and Yunnan Observatories for their valuable assistance. This work was supported by the National Astronomical Observatories Project of the Chinese Academy of Sciences (No. E4TQ2101), the National Natural Science Foundation of China (NSFC) under Grant No. 11427901, and the Chinese Meridian Project (CMP).

**References**

Babcock, H. W. 1953, ApJ, 118, 387 2